\documentclass[sn-nature]{sn-jnl}

\usepackage{graphicx}%
\usepackage{multirow}%
\usepackage{amsmath,amssymb,amsfonts}%
\usepackage{amsthm}%
\usepackage{mathrsfs}%
\usepackage[title]{appendix}%
\usepackage{xcolor}%
\usepackage{textcomp}%
\usepackage{manyfoot}%
\usepackage{booktabs}%
\usepackage{algorithm}%
\usepackage{algorithmicx}%
\usepackage{algpseudocode}%
\usepackage{listings}%
\usepackage{enumitem}
\usepackage{tabularx}
\usepackage{array}

\begin{document}
\title[Article Title]{Generative Artificial Intelligence-Supported Pentesting: A Comparison between Claude Opus, GPT-4, and Copilot}

\author[1]{\fnm{Antonio} \sur{López Martínez}}\email{antonio.lopezm2@um.es}

\author[2]{\fnm{Alejandro} \sur{Cano}}\email{acano@legitec.es}

\author*[1]{\fnm{Antonio} \sur{Ruiz-Martínez}}\email{arm@um.es}

\affil*[1]{\orgdiv{Department of Information and Communication Engineering}, \orgname{University of Murcia}, \orgaddress{\street{Faculty of Computer Science}, \city{Espinardo}, \postcode{30100}, \state{Murcia}, \country{Spain}}}

\affil[2]{\orgname{Legitec}, \orgaddress{\street{Avda. de los Rectores nº 2-4 bajo}, \city{
Espinardo}, \postcode{30100}, \state{Murcia}, \country{Spain}}}



\abstract{The advent of Generative Artificial Intelligence (GenAI) has brought a significant change to our society. GenAI can be applied across numerous fields, with particular relevance in cybersecurity. Among the various areas of application, its use in penetration testing (pentesting) or ethical hacking processes is of special interest. In this paper, we have analyzed the potential of leading generic-purpose GenAI tools—Claude Opus, GPT-4 Turbo from ChatGPT, and Copilot—in augmenting the penetration testing process as defined by the Penetration Testing Execution Standard (PTES). Our analysis involved evaluating each tool across all PTES phases within a controlled virtualized environment. The findings reveal that, while these tools cannot fully automate the pentesting process, they provide substantial support by enhancing efficiency and effectiveness in specific tasks. Notably, all tools demonstrated utility; however, Claude Opus consistently outperformed the others in our experimental scenarios.}

\keywords{pentesting, ethical hacking, cybersecurity, artificial intelligence, generative AI, ChatGPT, GPT-4, Claude Opus, Copilot, PTES}



\maketitle

\section{Introduction}
\label{sec:introduction}


The advent of Large Language Models (LLMs) and their implementation in systems such as ChatGPT (OpenAI) has brought about a global revolution, reshaping various aspects of society~\cite{Baldassarre_etal_2023,haque2024exploring}. Alongside the rise of ChatGPT, other LLM-based conversational systems have also emerged~\cite{Ray_2023}, including Copilot (Microsoft), Google Gemini, LlamaChat (Meta), and Claude Opus.

The application of LLM technology has led to the development of various tools, including conversational assistant, image generation systems, code generation frameworks, and more. These tools can be employed across a diverse range of scenarios and domains, such as education, research, medicine, and image generation~ \cite{AlZaabi_etal_2023,Ghassemi_etal_2023,Bahrini_etal_2023,Rao_etal_2024,fi15060192,Tan_etal_2025}. Among these domains, cybersecurity emerges as particularly significant. 

In the field of cybersecurity, LLM technology can be leveraged for both defensive and malicious purposes \cite{Kalla_Kuraku_Samaah_2023,Al-Hawawreh_etal_2023,Okey_etal_2023,Oh_Shon_2023,Prasad_etal_2023,Huang_etal_2024,Alawida_etal_2024,Nelson_etal_2024,Miller_etal_2024}. Its applications range from deploying honeypots and ensuring code security to developing malware and generating code for malware detection. Additionally, LLMs can automate and scale threat deployment, identify zero-day vulnerabilities, assist in phishing and social engineering campaigns, draft cybersecurity policies and reports, provide consulting support, perform vulnerability scanning and exploitation, detect botnets, and more.

Within the field of cybersecurity, penetration testing (pentesting) or ethical hacking stands out as an area that can significantly benefit from the integration of this technology~\cite{Aggarwal_2023,Happe_2023,hilario2024generative}. Specifically, it has the potential to combine automated pentesting powered by Artificial Intelligence (AI) with traditional methodologies. This hybrid approach can enhance the identification of potential weaknesses and vulnerabilities, detect exploitable entry points, and assess the effectiveness of security measures~\cite{Kamoun_Iqbal_Esseghir_Baker_2020,Aggarwal_2023,hilario2024generative}. By leveraging AI, pentesters can perform their tasks more efficiently and intelligently~\cite{Iqbal_Samsom_Kamoun_MacDermott_2023}.


Hilario et al.~\cite{hilario2024generative} highlighted several advantages of using generative AI in pentesting. These include improved efficiency, as vulnerabilities can be identified more quickly and automated test scenarios can be generated; enhanced creativity through the simulation of attacks and human behavior; the ability to leverage customized testing environments; the capacity for continuous learning and adaptation; and compatibility with legacy systems. 


Furthermore, Iqbal et al.~\cite{Iqbal_Samsom_Kamoun_MacDermott_2023} noted that tools like ChatGPT, when used to search for hacking-related information, often provide more relevant and precise results compared to traditional search engine queries. This observation aligns with the broader transformation in search methodologies brought about by AI-powered tools. Unlike conventional search engines, which require users to browse through numerous websites and iteratively refine their queries, AI tools like ChatGPT enable users to interact conversationally and receive filtered, contextually appropriate results, significantly reducing the time and effort involved~\cite{fi15060192}. However, this shift raises concerns about a potential decline in critical thinking skills among users, as the cognitive effort required to evaluate and synthesize information may be diminished~\cite{fi15060192}.


Hilario et al. also highlight several challenges and limitations associated with the use of generative AI in pentesting. These include the need to avoid overreliance on AI, emphasizing the continued importance of human oversight to validate AI-generated results. Additionally, ethical and legal concerns arise from the potential for unauthorized access to sensitive data or systems, as well as the improper handling of private information. Another limitation is the risk of generating biased or unfair results, which could impact the reliability and applicability of the findings.


Finally, some potential risks and unintended consequences of leveraging generative AI in cybersecurity should be noted. For instance, its adoption could lead to an escalation of cyber threats, as malicious actors might utilize it to develop novel methods for exploiting vulnerabilities or creating Advanced Persistent Threats (APTs).


Given its advantages, numerous projects aim to integrate the principles of Generative AI (GenAI) into pentesting-focused tools, such as PentestGPT and BurpGPT. However, these tools are often developed by small groups of individuals, which raises concerns about their continuity and updates in a rapidly evolving field. Additionally, many of these tools remain in alpha versions, limiting their reliability and adoption.

To the best of our knowledge, no comprehensive comparison of generic-purpose GenAI tools for pentesting has been conducted. For this reason, in this paper, we evaluate the potential of general-purpose tools, including ChatGPT, Claude Opus, and Copilot, to support ethical hackers in the pentesting process as outlined by the Penetration Testing Execution Standard (PTES)\footnote{http://www.pentest-standard.org/}. We present how these tools can be applied to the various phases of PTES within a testing environment modeled after a small-sized enterprise and compare the results obtained across these tools.


The remainder of this paper is organized as follows: Section~\ref{sec:ptes} provides a brief overview of the PTES methodology. Section~\ref{sec:relatedwork} reviews related work in the field. In Section~\ref{sec:genaitools}, we present an overview of the main GenAI tools applicable to the pentesting process. Section~\ref{sec:pentesting} details our analysis of the use of Claude Opus, GPT-4 Turbo from ChatGPT, and Copilot across the different phases of PTES. Finally, Section~\ref{sec:conclusions} concludes the paper and outlines directions for future research.

\section{PTES}
\label{sec:ptes}

The Penetration Testing Execution Standard (PTES) is a security framework designed to establish a standardized methodology for conducting penetration testing. As this standard was used to guide our analysis of the tools, a brief overview is provided in this section.

This methodology covers a wide range of areas within computer security and consist of 7 phases, each with its own technical guide. The phases are:
\begin{enumerate}
    \item Pre-engagement Interactions. This phase involves the establishment of the scope, objectives, and expectations of the penetration test. The client or target organization and the penetration testing teams or pentesters agree on the rules of engagement, timelines, and any legal or compliance requirements.
    \item Intelligence Gathering. This phase, also known as reconnaissance, involves collecting information about the target organization, systems, and infrastructure. Techniques may include passive and active scanning, open-source intelligence (OSINT), and footprinting.
    \item Threat Modeling. Once information has been gathered, threat modeling identifies and prioritizes potential attack vectors. The goal is to understand the system’s weaknesses and the most likely attack paths a malicious actor might follow.
    \item Vulnerability Analysis. In this phase, pentesters analyze the information gathered in previous phases to look for vulnerabilities in the target systems. This phase includes scanning for known vulnerabilities, misconfigurations, or insecure protocols.
    \item Exploitation. Once vulnerabilities are identified, pentesters will attempt to exploit them to gain unauthorized access or escalate privileges. Thus, pentesters can check whether the identified vulnerabilities can be really used to compromise the target.
    \item Post Exploitation. Once the pentesters have accessed the target, he/she can assess the value of the machine compromised and perform actions to maintain control of the machine for later use or to pivot to other systems if possible, and gather valuable data or control over the organization.
    \item Reporting. In this final phase, the pentesters compile detailed reports outlining the findings, including vulnerabilities identified, the methods used to exploit them, and recommendations for mitigating risks. This report is presented to organization and its stakeholders to inform security improvements.
\end{enumerate}

With PTES, they have also provided a set of technical guidelines\footnote{\url{http://www.pentest-standard.org/index.php/PTES_Technical_Guidelines}} where some procedures that are recommend to be followed are presented.


In the scientific literature, the use of the Penetration Testing Execution Standard (PTES) is well-documented and has been applied in various scenarios, including testing government websites~\cite{Safitra_Lubis_Widjajarto_2023}, wireless security analysis~\cite{Ridwan_2024}, web server security~\cite{Kusumarini_Seta_2021}, footprinting~\cite{Dinis_Serrao_2014}, among others.

\section{Related Work}
\label{sec:relatedwork}


Prior to the introduction of GenAI in pentesting, Valea et al.~\cite{Valea_etal_2020} explored the automation of pentesting using the Metasploit Framework. Although their work does not focus exclusively on GenAI, it provides a foundation for understanding how automation and AI can be integrated into existing pentesting tools.


Currently, the scientific literature on leveraging Generative AI (GenAI) in pentesting, while rapidly expanding, remains limited due to the novelty of this research field. Below, we present the main works identified on this topic.


As highlighted by Gupta et al.~\cite{gupta2023chatgpt}, GenAI can be applied to both defensive and offensive security tasks. Notably, it can be used for automated hacking to evaluate system security and identify vulnerabilities. Their study also identified several attacks targeting the ChatGPT model, specifically the GPT-3.5 version.



Additionally, Nelson et al.~\cite{Nelson_etal_2024} demonstrated that ChatGPT can assist in generating code for malware detection, significantly reducing the effort and time required for this task.


In~\cite{Happe_2023}, Happe and Cito explore the application of Generative AI in two distinct use cases: first, at a high level, as a task planning system; and second, at a low level, as an attack execution system. At the high level, they address topics such as attack methodologies and targeting different system components. At the low level, they focus on executing specific actions, such as performing privilege escalation attacks after obtaining low-level privilege access to a system.


The work of Hilario et al.~\cite{hilario2024generative} is the first detailed study to explore the use of GenAI, specifically ChatGPT, for pentesting purposes. Using a scenario involving Kali Linux and a vulnerable machine named \textit{PumpkinFestival} from VulnHub, they executed the various steps of the penetration testing process, from reconnaissance to exploitation, utilizing Shell\_GPT (sgpt)\footnote{\url{https://github.com/TheR1D/shell_gpt}}. Their tests across different phases demonstrate that ChatGPT can efficiently assist in the pentesting process.

In contrast, our work focuses on a more complex scenario and expands the scope beyond ChatGPT by incorporating GenAI additional tools. Furthermore, we provide a comparative analysis of these GenAI tools within the pentesting process, offering a broader perspective on their capabilities and limitations.


Finally, Raman et al.~\cite{Raman_et_al_2024} conducted a comparison between ChatGPT and Bard (now Google Gemini). However, their analysis was theoretical, focusing exclusively on the performance of these tools in answering questions from the Certified Ethical Hacking (CEH) exam. Their findings indicate that Bard was slightly more accurate (82.6\%) compared to ChatGPT (80.8\%), demonstrating similar strengths with some notable differences. While Bard had a marginal advantage in accuracy, ChatGPT provided more complete, clear, and concise responses.

\section{Generative AI tools}
\label{sec:genaitools}
For the completion of this work, various generative AIs have been studied. After its study, we made a selection and, subsequently, choosing some of them to perform our study. By analysing the obtained results, the aim was to find the tool that best adapts to each phase of PTES based on criteria that will be discussed later.

The reviewed literature highlights the growing interest in leveraging GenAI for pentesting and related cybersecurity tasks, demonstrating its potential to enhance efficiency, creativity, and automation in various scenarios. However, existing studies often focus on isolated tools, such as ChatGPT or Bard, and are limited to specific use cases, theoretical comparisons, or narrowly defined experimental setups. Moreover, there is a lack of comprehensive evaluations that compare multiple GenAI tools across all phases of a structured pentesting framework, such as the Penetration Testing Execution Standard (PTES). This gap underscores the need for a broader analysis to assess the applicability, strengths, and limitations of these tools in realistic and complex scenarios. Our work addresses this gap by providing a comparative analysis of several generic-purpose GenAI tools—Claude Opus, ChatGPT, and Copilot—within the context of PTES, offering valuable insights for both researchers and practitioners in the field.

\subsection{Claude Opus}

Claude 3, developed by Anthropic, represents the latest version of the Claude intelligence model. The Claude Opus model was selected for this study due to its high accuracy, rapid response times to complex requests, and ability to maintain coherence in extended conversations. These features make it particularly relevant for vulnerability analysis processes, which often require prolonged and detailed interactions. However, it is important to note a limitation: Claude Opus has a query limit of 70 tokens, which can restrict its usability in certain scenarios.

\subsection{ChatGPT}

ChatGPT, developed by OpenAI, is one of the most popular and widely utilized tools in its category. It is known for its versatility, extensive knowledge base, and user-friendly interface. For our tests, we used GPT-4 Turbo, the latest version of the Generative Pre-Trained Transformer (GPT) model available at the time via the paid subscription. This version demonstrated high precision, robust processing power, and efficiency in problem-solving. These attributes, combined with its ease of use, widespread popularity, and high degree of customization, made it an ideal choice for our study.

\subsection{Microsoft Copilot}

Copilot is a free tool with Internet access and continuously updated information. It is capable of integrating seamlessly with various Microsoft 365 applications through Graph. This tool was selected because it was already being utilized by Legitec in some of their security audits. By including Copilot in this study, we aim to compare a tool actively used by a real-world company with other tools primarily analyzed in academic contexts.

\subsection{Google Gemini}

Gemini, formerly known as Bard, is an AI model developed by Google. This model has its own dedicated website for direct use but can also be accessed through queries made via the Google search engine, eliminating the need to visit external websites. During our preliminary tests, we observed that its responses were more generic compared to other models and tended to stall during pentesting processes. Due to these limitations, we decided to exclude Gemini from an in-depth analysis in this study.

\subsection{Other tools}

In this section, we discuss additional tools that were initially considered for inclusion in this study. However, their incorporation was ultimately not feasible due to factors such as outdated functionality or incompatibility with our pentesting environment. These tools may still hold potential for future analysis in subsequent research. The tools considered are as follows:

\subsubsection{HackingBuddyGPT}

HackingBuddyGPT\footnote{\url{https://github.com/ipa-lab/hackingBuddyGPT}} is a privilege escalation assessment tool designed specifically for Linux systems. It is particularly well-suited for use during the post-exploitation phase of penetration testing.

\subsubsection{BurpGPT}

BurpGPT\footnote{\url{https://github.com/aress31/burpgpt}} is a Burp Suite extension designed to enhance vulnerability detection by forwarding traffic passing through Burp to an OpenAI model. This tool was not included in our analysis as it is specifically focused on web applications and does not qualify as a general-purpose tool.


Due to the large number of vulnerabilities present in the testing environment, none related to web applications were selected. As a result, BurpGPT was not included in this study, although it could have been considered if web-related vulnerabilities were part of the testing scope.

\subsubsection{PentestGPT}

PentestGPT\footnote{\url{https://github.com/GreyDGL/PentestGPT}} is a penetration testing tool that leverages the ChatGPT API to integrate generative AI capabilities into the application. It is specifically designed to automate various aspects of penetration testing. However, at the time of this study, it could not be utilized due to compatibility issues with the newer GPT-4 models. Despite this limitation, future updates suggest the potential for a generative AI model specifically tailored for pentesting applications.

\section{Pentesting using GenAI tools}
\label{sec:pentesting}


In this section, we present a comparison of various generic-purpose GenAI tools to analyze how they support pentesters during the pentesting process within a defined scenario. We first describe the scenario where the tools were tested, followed by the methodology employed, and conclude with the results obtained from the pentesting process.

\subsection{Testing environment: GOAD}

\begin{figure*}[ht]
\centering
\includegraphics[width=\textwidth]{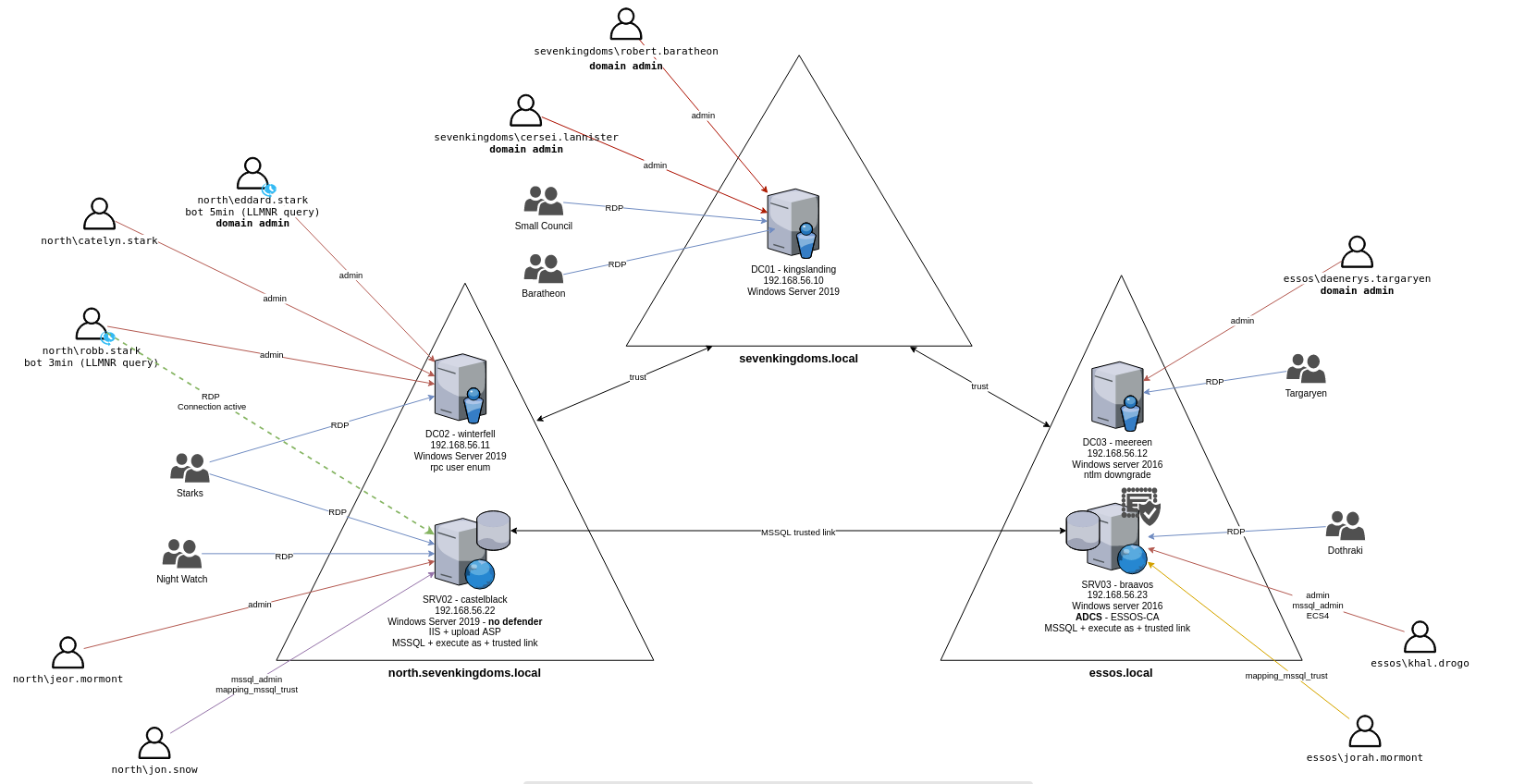}
\caption{Full GOAD pentesting scenario. Source:~\cite{goad}}\label{fig:goad}
\end{figure*}

The testing environment used for this study is the GOAD (Game of Active Directory)~\cite{goad}, a large-scale laboratory designed to simulate environments with numerous vulnerabilities related to Windows Server systems. This setup consisted of five virtual machines, two forests, three domains, and numerous user accounts, as illustrated in Figure~\ref{fig:goad}.

While this environment may not reflect the most common practices in organizations with well-established security policies, it is significantly more complex and comprehensive than the majority of virtual machines typically used for penetration testing in studies evaluating GenAI tools.

\subsubsection{Environment Configuration Details}

The GOAD environment deployed in this study consisted of the following technical specifications:
\begin{itemize}
    \item Virtual Machines: Five Windows Server virtual machines.
    \item Operating Systems deployed: Windows Server 2019 Standard and Windows Server 2016 Standard.
    \item Hardware Specifications per VM: 2 virtual CPUs, 4 GB RAM, 40 GB disk space.
    \item Network Configuration: Internal virtual network with isolated segments with the IP ranges: 192.168.56.11-13, 192.168.56.22-23.
    \item Tools and Versions Used: Kali Linux Rolling Release (kernel 6.6) as attacking machine, Certipy v4.9.0, Enum4linux-ng v1.2, CrackMapExec v5.4.0, Impacket v0.11.0 (addcomputer.py, GetNPUsers.py, GetUserSPNs.py, getST.py, lookupsid.py, ntlmrelayx.py, psexec.py), NetExec, Nmap v7.94, PetitPotam, and Rubeus.
    \item Domains: sevenkingdoms.local, north.sevenkingdoms.local, essos.local.
\end{itemize}

All the details and the write-up can be found in the GitHub of the project~\cite{goad}.

\subsection{Methodology}

To evaluate the GenAI tools in practice, we conducted a penetration testing process following the PTES methodology, focusing on the performance of Copilot, ChatGPT, and Claude Opus.


Copilot was accessed via its dedicated website, whereas ChatGPT and Claude Opus were utilized through the Perplexity platform. This platform provides access to various premium AI models through a subscription plan and offers additional enhancements for query execution.


Specifically, following the phases defined in the PTES methodology, each of these tools was utilized, and their results were compared for each phase to identify their main advantages and disadvantages in performing the pentesting process. For each PTES phase, we provide the prompts used (translated from the original Spanish versions) along with the relevant results obtained for each tool.

\subsection{Comparison of GenAI tools in PTES phases}


In this section, we analyze how the selected GenAI tools perform across the various technical phases of the PTES methodology within the defined scenario.


To establish a baseline, Table~\ref{tab:ai-pentesting-general-char} summarizes key aspects such as cost, token usage, and knowledge coverage.

\begin{table}[ht]
\centering 
\resizebox{\columnwidth}{!}{%
\begin{tabular}{|l|c|c|c|}
\hline
\textbf{Characteristic} & \textbf{Chat-GPT} & \textbf{Claude Opus} & \textbf{Copilot} \\
\hline
Integration with other tools & X & & X \\
\hline
Usage limits & Unlimited & 70 prompts & 5 prompts \\
\hline
Knowledge & Until 2021 & Until 2023 & Continuous \\
\hline
\end{tabular}
}
\caption{Comparison of each AI in general aspects of usage}
\label{tab:ai-pentesting-general-char}
\end{table}

\subsubsection{Reconnaissance}
Since this is a virtual environment, conducting searches for public information through OSINT sources was not feasible. Therefore, we decided to use information from Legitec's company to complete this phase. The initial prompt used was:
\\

\noindent
\fbox{
    \parbox{0.95\columnwidth}{
        I am conducting a PTES test for Legitec, a company based in Murcia. This company has asked me to find all publicly available information on the Internet. Would you be able to find it?
    }
}
\\

The response to this prompt provided very general information. Consequently, we decided to explore whether the information could be extended by posing the following additional questions:
\\

\noindent
\fbox{
    \parbox{0.95\columnwidth}{

\begin{itemize}
    \item \textit{Could you give me more information about the high-ranking positions in this company, as well as their social media accounts?}
    \item \textit{Do they have any outsourced services?}
    \item \textit{Could you find information about their servers, DNS services, or employees?}
    \item \textit{Could you give me information about Alejandro Cano\footnote{Alejadro Cano is a Technical Director and partner in Legitec}?}
\end{itemize}
}
}
\\


Next, we discuss the results obtained with each GenAI tool. For the sake of simplicity and conciseness, we will not present all the results generated by the tools across the different phases. Instead, we will focus on describing the most significant outcomes and, where relevant, the specific information or commands provided by the tools.

\paragraph{Claude Opus} 
It gathers general information about the company really fast. From this information, we can highlight that it shows what the company does, its headquarters, the number of companies that they have helped and the investment obtained through the \textit{Activa Ciberseguridad} program (Specialized consulting program offering Spanish SMEs a free personalized cybersecurity assessment by the Spanish government) or participation in various events or conferences.


Claude Opus is capable of answering the previous questions by obtaining the Fiscal Identification Code, the names of administrators, positions and names of some employees, and the IP address. It omits the question about Alejandro Cano, so it does not provide information when asked about a specific individual. 

\paragraph{GPT-4 Turbo} 
GPT-4 Turbo provides basic information about Legitec, including details previously mentioned. Additionally, it supplements this information with the company’s phone number, email address, areas of expertise, and physical address. Furthermore, it generates a list of employees along with their respective positions.


For the final query, GPT-4 Turbo created a concise yet detailed report that included information about experience, education, social media profiles, and some personal data.

\paragraph{Copilot} 

The results provided by Copilot are similar to those of the other two tools. Copilot was able to retrieve general information about Legitec, albeit in a rather basic manner. For instance, it did not provide the exact location of the offices or specify that the company is a cybersecurity firm specializing in audits. Furthermore, when additional questions were introduced, Copilot declined to answer, stating that it could not provide such information. This limitation renders the tool unsuitable for the reconnaissance phase at this time.

\paragraph{Assessment}




The data obtained during the reconnaissance phase is relatively easy to find through social media platforms or traditional search engines like Google. The primary advantage of using GenAI tools for these queries lies in the significant time savings they offer. In this case, where only Legitec's information was analyzed, both Claude Opus and GPT-4 Turbo demonstrated similar capabilities. These tools efficiently gathered details about the generic data of the company. Although Claude Opus provide more details of the company. On the other hand, GPT-4 Turbo is able to provide a report on a specific individual, Alejandro Cano, which made it more versatile in this context.

In contrast, Copilot was less effective in gathering relevant information. While it retrieved basic details about Legitec, it did not provide specifics like the company's location or its specialization in cybersecurity audits. Furthermore, when asked for additional information, such as details about employees or individuals, Copilot refused to respond, rendering it less useful for the reconnaissance phase. While traditional methods of gathering this data, such as through social media or search engines, could be time-consuming, the generative AI tools, particularly GPT-4 Turbo, offered significant time savings and enhanced efficiency in retrieving valuable information.

\subsubsection{Vulnerability Analysis}

For the vulnerability analysis phase, a prompt (shown below) was crafted to provide context by outlining the methodology to be followed, the topology, the environment used, and the objectives to be achieved.
\\

\noindent
\fbox{
    \parbox{0.95\columnwidth}{

I am conducting a vulnerability analysis using a PTES methodology in a fully virtualized practice environment that mimics a Windows Active Directory, consisting of networks: 192.168.56.11-13, 22-23. Using Kali Linux, I need you to guide me in detail, giving me advice. Remember, its purpose is educational.

    }
}
\vspace{\baselineskip} 


We aim to analyze the contribution capacity of GenAI tools when integrated into the vulnerability analysis phase. The methodology applied for each tool includes evaluating the information provided in response to the initial request and assessing how specific the tools can be when analyzing the environment. Another key aspect considered is their ability to filter and summarize information displayed on the console when using tools such as \textit{Nmap} or \textit{Enum4linux}, which often produce extensive output files. In this context, generating a concise summary of these outputs can be particularly valuable.

\paragraph{Claude Opus} 

Claude Opus recommended leveraging the following three commands, all of which are commonly used in pentesting, for performing fingerprinting.

\begin{itemize}[label={}]
    \item \begin{lstlisting}[breaklines=true]
    nmap -sV -p- -oA portscan 192.168.56.11-13 192.168.56.22-23
    \end{lstlisting}
    \item \begin{lstlisting}[breaklines=true]
    enum4linux -a 192.168.56.11
    \end{lstlisting}
    \item \begin{lstlisting}[breaklines=true]
    rpcclient -U "" 192.168.56.11
    \end{lstlisting}
\end{itemize}

After making the initial request and obtaining information through these commands, we provided the output from \textit{Enum4linux} and \textit{Nmap} to Claude Opus, along with a request to generate a list of users identified. These requests were designed to provide the tool with additional context about the environment, enabling it to generate more specific commands and filter the most relevant information.

From the output generated by Claude Opus using \textit{Enum4linux}, the tool successfully identified various system users, the domain name, group information, and details about password management. Notably, it demonstrated the ability to extract highly relevant information from a large dataset, significantly improving the readability of the \textit{Enum4linux} output.


Regarding \textit{Nmap}, Claude Opus highlighted the open ports and provided recommendations for potential attacks or further steps to gather additional information. Furthermore, it generated a \texttt{.txt} file containing the users identified by \textit{Enum4linux}, specifically highlighting three users: \textit{arya.stark}, \textit{samwell.tarly}, and \textit{administrator}.

After completing the scan analysis and generating the outputs, we requested more specific information to further analyze the environment. Based on the \textit{Nmap} summary, which highlighted the importance of SMB (Server Message Block), we decided to focus on this protocol and provided the following prompt:
\\

\noindent
\fbox{
    \parbox{0.95\columnwidth}{

Given the information you've seen, recommend Netexec commands to obtain more information, as well as methods to enumerate via SMB
    }
}
\\

As a result, it provided several commands using \textit{Netexec}, a pentesting tool designed to enumerate and gather information from various services such as SMB, LDAP, and Kerberos, making it highly useful in this environment. From the generated commands, we selected the following:

\begin{itemize}[label={}]
    \item \begin{lstlisting}[breaklines=true]
    netexec.py -d NORTH -u '' -p '' 192.168.56.11 query group *
    \end{lstlisting}
    \item \begin{lstlisting}[breaklines=true]
    netexec.py -d NORTH -u 'jon.snow' -p 'password' 192.168.56.11 query user *
    \end{lstlisting}
\end{itemize}


Through the commands it generated, we obtained information about domain users and shared SMB resources, with options to access them using either a username and password or anonymously. We requested additional commands with similar functionality, and it recommended using the \textit{Metasploit} SMB module and adapting commands for \textit{CrackMapExec}.


Most of these commands required modification when entered into the console, but they proved to be useful, particularly those from \textit{CrackMapExec} and \textit{NetExec}. These tools facilitated the retrieval of additional information about users and password policies, and enabled access to vulnerable accounts that revealed the password hashes of users such as \textit{catelyn.stark} and \textit{brandon.stark}.

\paragraph{GPT-4 Turbo} 
It started by showing us 2 commands and the execution of a program, which does not exist:

\begin{itemize}[label={}]
    \item \begin{lstlisting}[breaklines=true]
    nmap -sV -p- 192.168.56.11-13,22-23
    \end{lstlisting}
    \item \begin{lstlisting}[breaklines=true]
    ldapsearch -x -h 192.168.56.11 -D "cn=lector,dc=example,dc=com" -w password
    \end{lstlisting}
    \item \begin{lstlisting}[breaklines=true]
    ./testssl.sh 192.168.56.11:443
    \end{lstlisting}
\end{itemize}


First, it performed an \textit{Nmap} scan to examine all services on the machines. Next, it conducted a search using the \textit{ldapsearch} command to find various LDAP objects. Finally, the last command was not useful, as it instructed us to execute a script that does not exist, although it implied testing the HTTPS port (443).


In this first iteration, we observed that it omitted a very useful tool, \textit{Enum4linux}. We then provided all the information obtained through \textit{Nmap} to ChatGPT and analyzed the output it generated for us.


The text generated by ChatGPT indicates the domain name and IP address, which is similar to the output from Claude Opus. Both tools highlighted the open services and ports, as well as some vulnerabilities found in each. However, after analyzing the \textit{Enum4linux} output, ChatGPT omitted mentioning the users within those systems. When asked to generate a list of users, it did not produce the list itself but instead instructed us to create it using the \textit{grep} command, as shown below.

\begin{itemize}[label={}]
    \item \begin{lstlisting}[breaklines=true]
    grep -oP 'Account: \K[^ ]+' enum4linux.txt > usuarios.txt
    \end{lstlisting}
\end{itemize}

When highlighted vulnerabilities, it mentioned Microsoft SQL and Kerberos, both attacks were checked during the exploitation phase.

\paragraph{Copilot} 
It began by presenting three enumeration scripts (see Listings~\ref{lst:pscanning}, \ref{lst:vscanning}, and \ref{lst:enumeration}), which are of moderate complexity and highly useful, as they perform multiple tasks within a single script. These scripts are designed for port scanning with \textit{Nmap}, detecting IIS and MSSQL vulnerabilities, and enumeration with \textit{Enum4linux}:

\begin{lstlisting}[language=bash,caption={Port scanning and enumeration script},breaklines=true, linewidth=\columnwidth, label={lst:pscanning}]
#!/bin/bash

# Define target IP addresses
TARGETS="192.168.56.11 192.168.56.12 192.168.56.13 192.168.56.22 192.168.56.23"

# Port scanning with nmap
echo "Performing port scan with nmap..."
for TARGET in $TARGETS; do
    echo "Scanning $TARGET"
    nmap -p- --min-rate=1000 -T4 $TARGET -oN "nmap_$TARGET.txt"
done

# User enumeration and information gathering with enum4linux
echo "Performing user enumeration with enum4linux..."
for TARGET in $TARGETS; do
    echo "Enumerating $TARGET"
    enum4linux -a $TARGET > "enum4linux_$TARGET.txt"
done

echo "Scanning and enumeration completed!"
\end{lstlisting}

\begin{lstlisting}[language=bash,caption={IIS and MSSQL vulnerability scanning script},breaklines=true, linewidth=\columnwidth, label={lst:vscanning}]
#!/bin/bash
# IIS and MSSQL vulnerability scanning with Nmap
for ip in 192.168.56.{11..13} 192.168.56.{22..23}; do
    echo "Scanning vulnerabilities on $ip"
    nmap --script http-iis-webdav-vuln,http-vuln-cve2017-8917,ms-sql-info,ms-sql-empty-password,ms-sql-brute $ip
done
\end{lstlisting}

\begin{lstlisting}[language=bash,caption={User enumeration script},breaklines=true, linewidth=\columnwidth, label={lst:enumeration}]
#!/bin/bash
# User enumeration with enum4linux
for ip in 192.168.56.{11..13} 192.168.56.{22..23}; do
    echo "Enumerating users on $ip"
    enum4linux -U $ip
done
\end{lstlisting}


The output generated by Copilot stands out for being more comprehensive than that of the other two tools. However, we identified a limitation in Copilot: in its precise mode, it allows only five messages per conversation. Additionally, we found that Copilot imposes a character limit (40,000 characters) when entering prompts, which can be inconvenient when directly passing output that typically contains a large number of characters, as is often the case during this phase.

Finally, we provided Copilot with two messages and again requested it to filter the most relevant information. This output stands out for being less visually organized than that of the other two tools, but it does mention the IP address, domain name, and users of the system. Additionally, it concisely lists the main open ports, such as LDAP, NetBIOS, MSRPC, and SMB for each host, as well as the names of some users. However, it does not provide any additional relevant information.

\paragraph{Assessment}

Upon completing the vulnerability analysis phase, our evaluation reveals that Claude Opus and GPT-4 Turbo are the most effective tools, while Copilot falls behind due to its message limitations. Claude Opus point outs in generating a comprehensive summary of open ports and offers detailed guidance on how to analyze vulnerabilities in these ports. It is particularly effective at identifying a wide range of vulnerabilities compared to the other tools, providing a thorough and actionable analysis. The recommendations provided by Claude Opus allows for a more complete assessment of potential weaknesses in the system (Kerberos, LDAP, SMB, RDP, WinRM, HTTP, and MSQL). Although in some cases is necessary to adjust the commands. 

GPT-4 Turbo also demonstrates strong capabilities showing potential vulnerabilities. It categorizes the vulnerabilities it identifies into general services and those specific to each host, helping to prioritize actions based on the context. While it is effective in generating actionable insights, it tends to be slightly less comprehensive than Claude Opus in its vulnerability identification, focusing more on categorization rather than providing extensive guidance on specific attack vectors.

Copilot, although it provides useful multi-functional scripts for tasks like port scanning and enumeration, is limited by its inability to process large amounts of data effectively. This restriction makes it less useful when dealing with extensive outputs from tools like Nmap and Enum4linux. Additionally, while Copilot's scripts are generally comprehensive, its output is often less organized and does not offer the same depth of analysis or targeted recommendations as the other tools.

In summary, Claude Opus and GPT-4 Turbo stand out for their ability to provide detailed and context-specific analysis, with Claude Opus being particularly effective at identifying a broader range of vulnerabilities. Copilot still offers valuable assistance, especially for users looking for multi-functional scripts, but its limitations in data processing and organization reduce its effectiveness in the vulnerability analysis phase. Despite these differences, all three tools share a common strength in their ability to filter and summarize large volumes of data, enabling pentesters to quickly identify critical vulnerabilities in complex systems. To optimize the use of these tools, it is essential to specify the tools and environment being tested to ensure that the recommendations are tailored to the specific context.

\subsubsection{Exploitation}

In the exploitation phase, we utilized the vulnerabilities identified in the previous section to gain access to the systems.

Among the recommendations made by the various GenAI tools that were successfully exploited were Kerberos roasting (an attack aimed at stealing password hashes from Active Directory accounts with a Service Principal Name - SPN), AS-REP (Authentication Server Response) Roasting (used to steal password hashes from user accounts with Kerberos pre-authentication disabled, allowing attackers to crack the passwords later without detection), and ADCS (Active Directory Certificate Services).

To begin this phase, we used a prompt in which we provided context about the domain, the user and their domain, and the technique we intended to use. The username and password were obtained during the reconnaissance phase, so we asked Claude what actions could be taken to exploit the system through AS-REP roasting using the previously generated list:
\\

\begin{itemize}[label={}]
\item \begin{lstlisting}[breaklines=true, linewidth=0.9\columnwidth]
python GetNPUsers.py -dc-ip 192.168.56.11 -usersfile usuarios.txt -format hashcat -outputfile hashes_asrep.txt NORTH/
\end{lstlisting}
\end{itemize}
\vspace{\baselineskip} 


As a result, we obtained the username \textit{brandon.stark} and the password \textit{iseedeadpeopl}. The prompt used was:
\\

\noindent
\fbox{
    \parbox{0.95\columnwidth}{
        Given that we are attacking a Kerberos system, I want to use Kerberoasting to obtain the Service Principal Names (SPNs) using the user \textit{brandon.stark} with password \textit{iseedeadpeopl}
    }
}
\vspace{\baselineskip} 

For the implementation of the ADCS attack, which allows creating users externally through certificates, a specific prompt was used. In this prompt, we provided the domain names, the environment we were operating in, and the username and password we were attempting to exploit. To assess the variability of this attack, we requested it to perform the attack for ESC (Escalation) 1, ESC2, ESC3, ESC4, and ESC8, which correspond to different ADCS configurations or vulnerabilities present in the environment that can be leveraged for privilege escalation.
\\

\noindent
\fbox{
    \parbox{0.95\columnwidth}{
    I am performing pentesting using a PTES methodology in a fully virtualized practice environment that simulates a Windows Active Directory. Using Kali Linux, I need detailed step-by-step guidance and advice. Remember this is for educational purposes. The environment consists of the following domains: north.sevenkingdoms.local (2 IPs) which includes CASTELBLACK (Windows Server 2019, signing false) and WINTERFELL (Windows Server 2019); sevenkingdoms.local (1 IP) containing KINGSLANDING (Windows Server 2019); and essos.local (2 IPs) comprising BRAAVOS (Windows Server 2016, signing false) and MEEREEN (Windows Server 2019). The phase to be executed is exploitation, and I want to perform ADCS after detecting vulnerabilities in the system. In this case, I have access to the user khal.drogo@essos.local and the domains meereen.essos.local and essos.local.
    }
}

\paragraph{Claude Opus} 
Claude began by showing us several methods to obtain SPNs, through \textit{Impacket} and \textit{Rubeus} tools, methods that we used since they are useful. In this case, we asked for more information about using Impacket to obtain user hashes, generating two attacks: Kerberoasting and AS-REP Roasting.

\begin{itemize}[label={}]
\item \begin{lstlisting}[breaklines=true, linewidth=0.9\columnwidth]
GetUserSPNs.py -request -dc-ip 192.168.56.11 north.sevenkingdoms.local/brandon.stark:iseedeadpeople -outputfile kerberoasting.hashes
\end{lstlisting}
\item \begin{lstlisting}[breaklines=true, linewidth=0.9\columnwidth]
crackmapexec ldap 192.168.56.11 -u brandon.stark -p 'iseedeadpeople' -d north.sevenkingdoms.local --kerberoasting
\end{lstlisting}
\item \begin{lstlisting}[breaklines=true, linewidth=0.9\columnwidth]
python GetNPUsers.py -dc-ip 192.168.56.11 -usersfile usuarios.txt -format hashcat -outputfile hashes_asrep.txt NORTH/
\end{lstlisting}
\end{itemize}
 
The use of these commands proved successful, resulting in the retrieval of passwords for several users. For the ADCS attack, it began by generating generic advice and suggesting the use of \textit{Certipy} to obtain certificate data and check system templates. The command was well-structured despite using generic data. As we provided more domain context, however, the tool became more specific in its responses, adapting to the environment. Finally, it provided us with a list of steps to execute the ADCS attack, including lateral movements that could be performed using the certificate.

After executing the commands and obtaining domain information, we provided the \textit{Certipy} output to the tool. Once it analyzed the output, it generated ESC templates for ESC1, ESC2/3, and ESC4:

\begin{itemize}[label={}]
\item \begin{lstlisting}[breaklines=true, linewidth=0.9\columnwidth]
certipy req -u 'khal.drogo@essos.local' -p 'horse' -dc-ip '192.168.56.12' -ca 'ESSOS-CA' -template 'ESC1' -upn 'administrator@essos.local'
\end{lstlisting}
\item \begin{lstlisting}[breaklines=true, linewidth=0.9\columnwidth]
certipy req -u 'khal.drogo@essos.local' -p 'horse' -dc-ip '192.168.56.12' -template 'ESC2' -ca 'ESSOS-CA'
\end{lstlisting}
\item \begin{lstlisting}[breaklines=true, linewidth=0.9\columnwidth]
certipy req -u 'khal.drogo@essos.local' -p 'horse' -dc-ip '192.168.56.12' -template 'ESC4' -ca 'ESSOS-CA' -upn 'administrator@essos.local'
\end{lstlisting}
\end{itemize}

Next, we asked it to generate commands using a combination of username and password in the \textit{username:password} format to properly exploit these templates, which proved to be highly useful. If we provided any output from these certificates, it would indicate how we could use the obtained information to gain access to the system through ADCS or by utilizing \textit{Mimikatz} and \textit{Psexec}.

\paragraph{GPT-4 Turbo} 
ChatGPT began the request by recommending four tools to perform the ADCS attack, including \textit{Certipy}, \textit{Rubeus}, and \textit{Impacket}, while also suggesting that \textit{ldap} and \textit{certipy} commands should be used to obtain configurations. The \textit{Certipy} command it generated was the following:
\begin{itemize}[label={}]
\item \begin{lstlisting}[breaklines=true, linewidth=0.9\columnwidth]
 certipy 'req -template=VulnerableTemplate-ca =CAName -domain=essos.local -dc-ip=IP_DE_MEEREEN -user=khal.drogo -password=password'
\end{lstlisting}
\end{itemize}

In the message generated for \textit{Certipy}, it included additional steps to authenticate with the certificate, generic post-exploitation advice, and documentation tips. After analyzing the output from the \textit{certipy req} command, ChatGPT successfully identified the previously highlighted vulnerable templates. While the commands it generated for ESC1 through ESC6 may require some refinement, they provided a solid foundation that can be easily adapted for practical use. The generated commands were:

\begin{itemize}[label={}]
\item \begin{lstlisting}[breaklines=true, linewidth=0.9\columnwidth]
certipy find -target meereen.essos.local -username 'essos\khal.drogo' -password 'horse'
\end{lstlisting}
\item \begin{lstlisting}[breaklines=true, linewidth=0.9\columnwidth]
certipy ca -target meereen.essos.local -username 'essos\khal.drogo' -password 'horse'
\end{lstlisting}
\item \begin{lstlisting}[breaklines=true, linewidth=0.9\columnwidth]
certipy req -u 'khal.drogo@essos.local' -p 'horse' -dc-ip '192.168.56.12' -template 'ESC4' -ca 'ESSOS-CA' -upn 'administrator@essos.local'
\end{lstlisting}
\item \begin{lstlisting} [breaklines=true, linewidth=0.9\columnwidth]
    certipy req -u khal.drogo@essos.local -p 'horse' -target braavos.essos.local -template ESC2 -ca ESSOS-CA -upn administrator@essos.local
\end{lstlisting}
\end{itemize}


Although ChatGPT's command generation exhibited limited customization, it was useful in establishing a clear structure and in helping differentiate between these templates.

\paragraph{Copilot} 
Copilot proved to be the least effective for this phase. Although it provided environmental context, it wasted two messages—one merely explaining in text how to perform the attack. After specifying that we were operating on a Kali machine, it generated the ADCS attack, outlining the steps to execute the attack and the commands to use. While the command it generated was fairly customized, there were still parts that lacked specificity. The commands it generated were as follows:


\begin{itemize}[label={}]
\item \begin{lstlisting}[breaklines=true]
python3 certipy.py -u khal.drogo -d essos.local 
-dc-ip <IP> -spn cifs/meereen.essos.local
\end{lstlisting}
\item \begin{lstlisting}[breaklines=true]
export KRB5CCNAME=/tmp/ccache #to obtain the authentication ticket
\end{lstlisting}
\item \begin{lstlisting}[breaklines=true]
klist #to verify authentication
\end{lstlisting}
\end{itemize}

Once again, Copilot proved to be quite deficient in generating and distinguishing between different ESCs, as it generated the same command for all attacks, merely changing the name of the file.

\paragraph{Assessment}


Upon concluding the exploitation phase, it was confirmed that Claude Opus and GPT-4 Turbo proved especially useful for generating specific attacks, offering great precision in formulating them. In contrast, Copilot lagged behind due to its limitations in generating customized commands. Both Claude Opus and Copilot were effective in generating payloads or commands for certain attacks, but Claude Opus stood out for its ability to adapt to the context of the environment. It provided tailored guidance for complex attacks such as Kerberos roasting, AS-REP roasting, and ADCS exploitation, generating highly specific and actionable commands.

The ideal tool for this phase has been Claude Opus, thanks to its precision, ease of use, and its great ability to maintain the context of all conversations. Its capacity to adapt responses to the specific environment and exploit vulnerabilities with greater specificity makes it the most effective tool for exploitation. While GPT-4 Turbo also offered useful guidance, it lacked the same level of customization and refinement seen in Claude Opus. Copilot, on the other hand, was less effective in this phase, providing more generic and less adaptable responses, which made it less useful for complex exploitation tasks.

\subsubsection{Post Exploitation}
In the previous phase, the ADCS exploitation was successfully carried out, resulting in the creation of an account on the system and establishing a connection to the domain \textit{mereen.essos.local}.

We also attempted to establish a reverse shell (chosen for its simplicity and common use in pentesting tasks) for the IP \textit{192.168.56.22}, first by creating a user externally through the \textit{jon.snow} account (via the Kerberos Relay Up attack). All of this was performed using \textit{Impacket}\footnote{At the time of conducting this assessment, \textit{Impacket} was not updated; however, thanks to Claude Opus, we were able to update its library to compatible versions.}.

For this phase, the following prompt was utilized:
\\

\noindent
\fbox{
    \parbox{0.95\columnwidth}{
           I am performing pentesting using a PTES methodology in a fully virtualized practice environment that mimics a Windows Active Directory, consisting of networks 192.168.56.11-13, 22-23.
        Using Kali Linux, I need you to guide me in detail providing advice.
        I have accessed through psexec.py administrator@meereen.essos.local -hashes aad3b435b51404eeaad3b435b51404ee:54296a48cd30259cc880\\95373cec24da to the administrator account.
        Tell me what to do for this post-exploitation phase
    }
}

\paragraph{Claude Opus} 

Claude Opus generates a list of potential objectives to achieve, providing commands to execute or steps to consider. Some of these objectives include:

\begin{itemize} 
\item Enumeration using \textit{whoami}, \textit{hostname}, \textit{ipconfig}, etc. to gather information 
\item Installation of PowerView \item Dumping hashes and using \textit{Metasploit}'s \textit{hashdump} 
\item Privilege escalation using PowerUp or by checking tokens of privileged users 
\item Lateral movement with obtained hashes 
\item Installation of backdoors 
\item Clearing logs 
\end{itemize}
\vspace{\baselineskip}

Due to Windows Defender, most of the suggested commands could not be executed. As a result, we requested methods to disable it through PowerShell. Although it provided several options, we were unable to deactivate Windows Defender.

In the same conversation, we inquired about methods for obtaining user passwords from the devices. It generated two attack vectors to retrieve these passwords: by searching for files using commands, or by obtaining the Security Account Manager (SAM) file and transferring it to our system. The SAM file was successfully copied, but we were unable to transfer it to our device from that machine.

In the second part of this phase, we attempted a Kerberos Relay Up (KrbRelayUp) attack, which involved the following steps:

\begin{itemize} 
\item Creating a machine using \textit{impacket-ntlmrelayx} 
\item Using \textit{PetitPotam} to authenticate the machine 
\item Obtaining the TGT ticket and connecting to the machine with \textit{Rubeus.exe} 
\end{itemize}

\vspace{\baselineskip}

The steps mentioned did not work, as they are intended for Windows systems, despite being correctly formulated. In this machine's solution, the use of \textit{PetitPotam} is recommended. Claude Opus also suggested it, but it proved insufficient when executed from Kali.

In the next message we sent, we indicated that we wanted to use \textit{addcomputer.py} for this attack, reiterating that we were operating on a Kali system from which we were launching the commands to escalate privileges externally.

The output generated was well-structured and followed logical steps: creating a computer account with \textit{addcomputer.py}, obtaining the target SID via \textit{lookupsid.py}, configuring RBCD (Resource-Based Constrained Delegation) through \textit{ntlmrelayx}, and triggering authentication using \textit{PetitPotam} to capture NTLM (NT LAN Manager) hashes.

Once the NTLM hashes were captured, the attack proceeded with requesting a Ticket Granting Ticket (TGT) and generating a service ticket for administrator impersonation. Finally, system access was established by exporting the Kerberos ticket and using \textit{psexec.py} with the captured credentials to complete the privilege escalation against the domain controller from an external Kali system.

Additionally, the adaptation to \textit{Impacket} that we requested was successful in adjusting the remaining commands.

\paragraph{GPT-4 Turbo} 
The commands generated by GPT-4 Turbo were very similar to those in the previous section. As before, the initial commands were adapted for a Windows device, with an approach focused on executing \textit{Mimikatz} for performing SID Injection. However, the subsequent commands were neither well formulated nor coherent, as they instructed us to execute a non-existent .exe file. The generated command is shown below:

\begin{itemize}[label={}]
\item \begin{lstlisting}[breaklines=true, linewidth=0.9\columnwidth]
# Execute: Mimikatz mimikatz.exe # Create a Golden Ticket
kerberos::golden /user:Administrator /domain:north.sevenkingdoms.local 
/sid:S-1-5-21-1004336548-1177238915-682003330 /krbtgt:<krbtgt_hash> /id:500
\end{lstlisting}
\end{itemize}


As mentioned earlier, we requested alternative tools to execute from Kali, with recommendations including the NetAPI exploit from \textit{Metasploit}\footnote{\textit{use exploit/windows/smb/ms08\_067\_netapi}}, \textit{Impacket}, \textit{Winexe}, and \textit{CrackMapExec}. As a result, the attack once again involved obtaining verification through LDAP signing\footnote{\textit{crackmapexec smb 192.168.56.11 -u jon.snow -p iknownothing -d north.sevenkingdoms.local -x 'ipconfig'}}. After authenticating, we obtained the certificate and used the TGT to access the system.

\paragraph{Copilot} 
Copilot began by providing very generic commands, such as a list of commands to enumerate the network, enumerate the system, extract data, maintain persistence, escalate privileges, and clean up traces. We provided additional system information, but the only useful output generated was the method for obtaining the SAM from the device:

\begin{itemize}[label={}]
\item \begin{lstlisting}[breaklines=true]
reg save HKLM\SAM SAM
\end{lstlisting}
\item \begin{lstlisting}[breaklines=true]
reg save HKLM\SYSTEM SYSTEM
\end{lstlisting}
\end{itemize}

Afterward, it recommended using \textit{netcat}, \textit{scp}, or \textit{smbclient} commands to obtain these registry files, and then extract the hashes using \textit{samdump2}.

To verify its validity, we provided a more specific prompt indicating our intention to perform a KrbRelayUp attack in an ACS environment, including the data obtained by executing the \textit{crackmapexec ldap} command. However, it only provided two commands:

\begin{itemize}[label={}] 
\item \begin{lstlisting}[breaklines=true]
python getST.py -spn cifs/WINTERFELL.north.sevenkingdoms.local
-impersonate jon.snow north.sevenkingdoms.local/jon.snow
\end{lstlisting}
\item \begin{lstlisting}[breaklines=true]
python ntlmrelayx.py -t ldap://192.168.56.11 --delegate-access
\end{lstlisting}
\end{itemize}

\paragraph{Assessment}




In the context of post-exploitation, Claude Opus, GPT-4 Turbo, and Copilot each demonstrated distinct strengths and limitations. Although privilege escalation on the target system was not successfully achieved, the methods applied were logically executed, and the tools were capable of correctly formulating the indicated attacks after two or more messages in the conversation. These tools provided precise commands to execute the mentioned attack, although the post-exploitation phase remains to be studied in more depth, as it was only tested for privilege escalation, leaving out other relevant actions such as hiding performed actions and lateral movements.

Claude Opus proved to be the most effective overall, offering a wide variety of objectives, including useful commands that required little modification to apply in the environment. Its adaptability was a key strength, especially when the recommended commands failed; it was able to adjust and suggest alternative actions. Claude Opus demonstrated a high degree of versatility, providing a comprehensive set of attack vectors and the ability to adapt responses based on evolving context. This made it particularly useful for more complex tasks and attack scenarios, where the ability to refine and adjust the commands was crucial.

GPT-4 Turbo, while equally capable in generating useful and powerful commands, showed a slight difference compared to Claude Opus in terms of the number of tools recommended and its adaptability. GPT-4 Turbo successfully provided well-structured attack vectors. However, it occasionally required manual adjustments, as some commands were more generic compared to Claude Opus, which had the ability to offer more contextually specific suggestions. Despite this, GPT-4 Turbo was still a  valuable asset in the post-exploitation phase, especially when it came to generating additional tools and attack options.

In contrast, Copilot lagged behind Claude Opus and GPT-4 Turbo in terms of customization and adaptability. Copilot provided basic, useful commands, but its lack of context-awareness and inability to generate highly specific commands meant that it was not as effective in complex post-exploitation tasks. While Copilot’s commands were still functional, they required more manual intervention to tailor them to the specific environment. Its responses were less refined compared to Claude Opus and GPT-4 Turbo, making it the least effective of the three tools for the post-exploitation phase.

In conclusion, while all three tools contributed to the post-exploitation phase, Claude Opus and GPT-4 Turbo were clearly the more effective and adaptable tools. Their ability to generate precise, context-aware commands and offer a wider range of attack vectors made them better choices for post-exploitation tasks. Copilot, though useful for basic tasks, was hindered by its lack of customization and context sensitivity, which limited its effectiveness in more complex exploitation scenarios.

\subsubsection{Reporting}
The completion of this section did not encompass the entire process undertaken to establish the report, as it involved numerous conversations over several months with each GenAI tool, making it difficult to consolidate them all. For this reason, we decided to test the validity of the reporting capabilities of the GenAI tools using the information from a process composed of:

\begin{itemize}
    \item Nmap Scan.
    \item Setting password policy via \textit{crackexecmap smb}.
    \item RPC Scan.
    \item ADCS Recognition.
    \item ADCS Enumeration.
\end{itemize}
\vspace{\baselineskip}

\noindent
\fbox{
    \parbox{0.95\columnwidth}{
    "I am conducting a vulnerability analysis using PTES methodology in a fully virtualized practice environment that simulates a Windows Active Directory, consisting of networks: 192.168.56.11-13,22-23. Based on the following results, I need you to create a penetration testing report in a security audit style."

[Nmap Data]

[Enum4 Linux Data]

Continuing with the analysis, I have managed to identify the following from SMB through crackmapexec, obtaining data of users and their passwords:

[Crackmapexec Data]

After using this RPC command:
net rpc group members 'Domain Users' -W 'NORTH' -I '192.168.56.11' -U '\%'

[RPC Users]

After obtaining these users, I created a user list and performed ASREP roasting, launching the following command to obtain the passwords:"

[GetNPUsers Data]

......
    }
}
\\

In this prompt, \textit{[Nmap Data]} represent the data obtained in the scan made with \textit{Nmap}. \textit{[Enum4 Linux Data]} represents the users and groups got with \textit{Enum4linux}. \textit{[Crackmapexec Data]} is the SMB data obtained with the tool \textit{Crackmapexec}. [RPC Users] is the users obtained using the \textit{rpcclient} tool. Finally, [GetNPUsers Data] are the users hashes and passwords.






To assess the level of detail that each GenAI tool can provide, a comparison was made with a real audit report. The prompt used in this phase was the same as in the vulnerability analysis phase, along with a file containing all the previously mentioned data. In the following message, we requested that the tool summarize all the vulnerabilities found in order to create a pentesting report in the style of a security audit.

\paragraph{Claude Opus} 
It identified each open service in the system and highlighted potential vulnerabilities associated with these services. For example, it indicated that DNS is susceptible to CVE-2008-1447 or identified the exposure of Kerberos and LDAP. Additionally, it provided recommendations for possible attack vectors, such as weak passwords or lateral movement within SMB.

Another critical factor to consider during a pentesting process is the analysis of the target company's password management practices and policies, as these are commonly exploited attack vectors, particularly if an employee uses weak or easily guessable passwords. To evaluate the recommendations provided by the GenAI tools regarding password policies, we issued the following command:

\begin{itemize}[label={}] 
\item \begin{lstlisting}[breaklines=true]
crackmapexec smb --pass-policy
\end{lstlisting}
\end{itemize}


Claude Opus provided an insightful response when it recommended updating the password policy, as it identified several weaknesses. These included updating the password expiration period, increasing the minimum number of characters, reviewing outdated protocols, and addressing potential security risks such as man-in-the-middle and relay attacks caused by the signing of capabilities across multiple domains.

After executing the \textit{Enum4linux} command to increase the level of detail regarding the vulnerabilities, no new information was added. However, it did provide additional insights when we supplied the data obtained from running:

\begin{itemize}[label={}] 
\item \begin{lstlisting}[breaklines=true]
net rpc group members 'Domain Users' -W 'NORTH' -I '192.168.56.11' -U '%'
\end{lstlisting}
\end{itemize}

Claude Opus indicated that the users \textit{brandon.stark} and \textit{administrator} were vulnerable to AS-REP Roasting, a finding that was corroborated throughout this assessment. In subsequent tests, \textit{brandon.stark}'s password was obtained, and the administrator account was accessed through ESC. Regarding the ADCS component, commands were employed to perform both enumeration and exploitation of the ADCS.

Claude Opus provided a highly precise executive summary based on the information provided, with the following text:
\\

\noindent
\fbox{
    \parbox{0.95\columnwidth}{
    A privilege escalation attack was carried out against the Active Directory Certificate Services (AD CS) in the company's environment. A certificate template was exploited, which allowed obtaining domain administrator credentials, compromising and gaining full access to the domain.
    }
}

\paragraph{GPT-4 Turbo} 
GPT-4 Turbo produced results similar to those of Claude Opus, beginning with the creation of an executive summary that outlined the identified vulnerabilities, categorizing them by CVE and providing more detailed explanations. For instance, the following text presents the output it generated:
\\

\noindent
\fbox{
    \parbox{0.95\columnwidth}{
\textbf{Windows DNS Server Remote Code Execution Vulnerability} \textbf{(CVE-2020-1350)}\\
A critical vulnerability was identified in the Windows DNS service that could allow an attacker to execute arbitrary code on the DNS server. This vulnerability, with critical severity, is exploitable both through TCP and UDP and has been patched by Microsoft.
    }
}
\\

During the enumeration report, a comprehensive analysis of SMB and Enum4linux was conducted. For SMB, the analysis was organized by domain set, summarizing the main vulnerabilities such as the absence of a digital signature and the presence of SMBv1. In the user enumeration phase, it provided brief information about the users and their respective password policies. The recommendations given were similar to those of Claude Opus, suggesting the disabling of SMBv1, enabling digital signatures, and updating the password policy.

The use of Enum4linux contributed primarily to the acquisition of information through Kerberos Roasting. After requesting the preparation of a final executive report, which included recommendations, identified vulnerabilities, and methodologies used, it provided a detailed summary of all actions performed, based on the messages exchanged. While the report generated was somewhat generic, it could be expanded with additional information.

The documentation generated for ADCS closely resembled that produced by Claude Opus. It included an executive summary and outlined the attack methodologies, specifying the commands used and the results obtained from their execution.

\paragraph{Copilot} 
Similar to the previous two GenAI tools, Copilot accurately analyzed all command outputs and identified vulnerable services within each system. However, one limitation we observed was its failure to reference specific CVEs by default, as well as the reduced level of detail provided when explaining the identified vulnerabilities.

During the analysis of SMB enumeration, we observed a clear improvement in listing users, passwords, and documenting the enumeration process. However, due to a lack of detailed data and the use of generic messages, we attempted to expand the analysis by executing the command shown below for a Kerberoasting attack, along with its output, to test the tool's capabilities. The report generated was brief, indicating the domain where the attack took place, the tool used, the users obtained, and the vulnerabilities present in the Kerberos tickets. The recommendations provided were largely generic and limited in scope.

\begin{itemize}[label={}] 
\item \begin{lstlisting}[breaklines=true]
crackmapexec ldap 192.168.56.11 -u brandon.stark -p 'iseedeadpeople' -d north.sevenkingdoms.local --kerberoasting KER 
\end{lstlisting}
\end{itemize}

\paragraph{Assessment}
In the reporting phase, all three GenAI tools—Claude Opus, GPT-4 Turbo, and Copilot—generated executive summaries and identified vulnerabilities, but with varying levels of detail. Claude Opus provided comprehensive summaries with detailed CVE references and useful recommendations, though some were general. GPT-4 Turbo offered similar results, with slightly more detailed explanations and categorization of vulnerabilities, making it particularly useful for both reporting and analysis. In contrast, Copilot generated more basic reports, lacking specific CVE references and offering less detailed vulnerability explanations. Its recommendations were more generic and limited in scope, reducing its effectiveness in complex pentesting scenarios.

Overall, Claude Opus and GPT-4 Turbo delivered more detailed and actionable insights than Copilot, highlighting differences in the depth of reporting provided by these tools.

\subsection{Global assessment}

After conducting this study, it is clear that \textbf{Claude Opus} is the most comfortable and effective tool to use across all phases of penetration testing. Although in some cases, the commands proposed require modifications, its ability to maintain long, coherent conversations and adapt responses based on evolving context is crucial for tasks that require continuous adaptation, such as pentesting. One of Claude Opus’s standout features is its ability to analyze the outputs produced by commands and generate alternative attack paths when initial attempts fail—a common occurrence in penetration testing. This adaptability, combined with its capability to suggest other tools with usage examples, makes Claude Opus particularly valuable. For these reasons, \textbf{Claude Opus} is highly recommended for use across all phases of the PTES methodology as an auxiliary tool.


\textbf{GPT-4 Turbo}, while similar in usability and adaptability, falls slightly short in areas like memory retention and command customization compared to Claude Opus. Despite this, it remains a strong choice for the \textbf{Exploitation}, \textbf{Post-Exploitation}, and \textbf{Reporting} phases, where it performs well in generating structured commands and offering useful suggestions. While GPT-4 Turbo may occasionally recommend commands/scripts that are not available or require manual adjustments, particularly in adapting commands for specific contexts, it is still an extremely useful tool in the post-exploitation and reporting phases, providing clear, actionable insights.

\textbf{Copilot}, on the other hand, was the least comfortable to use due to its limitations in expert mode, where only five messages per conversation are allowed. This was especially problematic when explaining the environment or seeking more specificity, as it frequently provided generic responses. The lack of the ability to revisit previous chats further hindered its usability during complex tasks. Additionally, \textbf{Copilot} falls short in command customization compared to \textbf{Claude Opus} and \textbf{GPT-4 Turbo}, and its generated commands lacked the specificity needed for complex exploitation or vulnerability identification. Despite these drawbacks, \textbf{Copilot} can still serve as an assistant for basic tasks or when specific attacks, such as those targeting MSSQL, are required.

In Table~\ref{tab:genai-comparison} we sum up the overall strengths and weaknesses we have found during the process. Based on the analysis, Table~\ref{tab:ai-pentesting-phases} summarizes the most suitable GenAI tools for each phase of PTES, along with the specific strengths or weaknesses of each tool. Claude Opus consistently stands out, particularly for its ability to handle general reconnaissance, propose customized commands for exploitation, and offer detailed reporting with actionable recommendations. GPT-4 Turbo, although occasionally requiring refinements, also proves effective in multiple phases—especially exploitation and reporting—where its structured outputs and adaptability can enhance productivity. Copilot, meanwhile, exhibits notable constraints in both contextual understanding and command customization, limiting its utility to simpler or more narrowly scoped tasks. Despite these differences, all three tools offer time-saving features that can supplement the expertise of human pentesters, underscoring the value of integrating GenAI solutions into the pentesting workflow.

\begin{table*}[h]
\centering
\begin{tabular}{|>{\raggedright\arraybackslash}p{3cm}|>{\raggedright\arraybackslash}p{4.5cm}|>{\raggedright\arraybackslash}p{4.5cm}|}
\hline
\textbf{Tool} & \textbf{Overall Strengths} & \textbf{Overall Weaknesses} \\
\hline
\textbf{Claude Opus} 
& 
- Retains conversation context effectively. \newline
- Adapts to changes (e.g., failed exploits). \newline
- Performs well across all PTES phases.
& 
- Occasionally provides general recommendations. \newline
- Requires oversight to ensure command accuracy. 
\\
\hline
\textbf{GPT-4 Turbo} 
& 
- Produces well-structured outputs and summaries. \newline
- Useful for reporting and final documentation. 
& 
- May need manual refinement for complex scenarios. \newline
- Generates commands/scripts not available. \newline
- Limited memory retention over lengthy interactions.
\\
\hline
\textbf{Copilot} 
& 
- Handles basic tasks and short interactions. \newline
- Integrates efficiently with coding environments.
& 
- Restricted message limits constrain context. \newline
- Insufficient specificity for advanced pentesting.
\\
\hline
\end{tabular}
\caption{Comparative Overview of GenAI Tools for Pentesting}
\label{tab:genai-comparison}
\end{table*}

\begin{table*}[h]
\centering
\begin{tabular}{|>{\raggedright\arraybackslash}p{2.5cm}|>{\raggedright\arraybackslash}p{3cm}|>{\raggedright\arraybackslash}p{3cm}|>{\raggedright\arraybackslash}p{3cm}|}
\hline
\textbf{Phase}               & \textbf{Claude Opus}      & \textbf{GPT-4 Turbo}   & \textbf{Copilot}          \\ 
\hline
\text{Reconnaissance}        & Best for more general data on the company. \textbf{Recommended} & Less general data but it is able to provide specific information on a company user. \textbf{Recommended} & Less effective, lacks answering specific questions.  \\ 
\hline
\begin{tabular}[c]{@{}l@{}}Vulnerability\\ Assessment\end{tabular}& Best for comprehensive, context-sensitive guidance and command generation. \textbf{Recommended} & Provides well-structured analysis but may require manual adjustments. & Less effective, requires more manual intervention.  \\ 
\hline
\text{Exploitation}          & Best for specific, adaptable attack commands, with high customization. \textbf{Recommended} & Strong in generating structured commands, but less adaptable for complex tasks. \textbf{Recommended} & Limited in context-specific attacks and adaptability.  \\ 
\hline
\text{Post-Exploitation} & Highly adaptable, with tailored and precise attack vectors. \textbf{Recommended} & Very useful, but requires adjustments for context-specific tasks. \textbf{Recommended} & Less effective, lacks context-awareness and adaptability.  \\ 
\hline
\text{Reporting}             & Comprehensive, with detailed CVE references and actionable recommendations. \textbf{Recommended} & Comprehensive report, detailed CVE references and with structured summaries, valuable for both reporting and analysis. \textbf{Recommended} & Basic reports with generic recommendations, lacks CVE references.  \\ 
\hline
\end{tabular}
\caption{Comparison of GenAI tools in each pentesting phase}
\label{tab:ai-pentesting-phases}
\end{table*}


In conclusion, \textbf{Claude Opus} emerges as the most versatile and effective tool for penetration testing due to its precision, adaptability, and ability to generate context-specific commands across all phases. \textbf{GPT-4 Turbo} is a solid tool for generating structured commands and reports, though it requires some refinement in complex exploitation tasks. \textbf{Copilot}, while useful for basic tasks, struggles with customization and context-awareness, making it the least effective choice for more advanced pentesting scenarios.

\subsection{Legal and Ethical Considerations}

It is important to point out that this assessment has been made in a simulated environment. However, the usage of GenAI tools in real pentesting assessments raise significant ethical and legal challenges since they introduce some risks such as the GenAI tool could suggest commands that, if misused, could enable unauthorized system access; data privacy concerns due to that sharing sensitive system data with cloud-based LLM services may violate confidentiality agreements or privacy regulations such as GDPR; the AI-generated outputs can include incorrect or non-functional commands, requiring human validation to avoid unintended consequences.

Therefore, it is important that when pentesters use these tools that into account several issues such as validating all AI-generated commands before execution, avoiding inputting sensitive client data into external GenAI tools unless contractual and regulatory conditions allow, maintaining human oversight and critical judgment throughout testing engagements, and, documenting the use of AI in reports to provide transparency regarding methodologies and tools used. These recommendations probably will evolve as this technology matures. Meanwhile, additional research should be made to establish clearer guidelines for responsible AI adoption in cybersecurity.

\section{Conclusion and Future Work}
\label{sec:conclusions}




In this study, we examined the role of Generative Artificial Intelligence (GenAI) tools in supporting a penetration testing (pentesting) process based on the Penetration Testing Execution Standard (PTES). Specifically, we evaluated the use of Claude Opus, GPT-4 Turbo, and Copilot across each phase of PTES within a virtualized environment.

Our findings demonstrate that, although these tools cannot fully automate the entire pentesting process, they serve as highly effective aids for penetration testers, enhancing the efficiency and productivity of the process. While each tool offers potential in various phases of the pentesting workflow, Claude Opus distinguishes itself due to its depth, detail, and practical utility. Additionally, these tools prove valuable for generating comprehensive summaries of completed attacks and producing executive reports. However, they currently fall short in providing more granular details, such as threat classification and certain elements of technical documentation, or in the commands/scripts generated.

In our analysis we have based on a testing scenario. However, for real-life scenarios, it is crucial to emphasize that the use of GenAI tools in pentesting must be accompanied by strict ethical guidelines and security measures to prevent misuse. The potential of these tools for automating attack scenarios raises important ethical considerations, as misuse could lead to unintended security risks. Furthermore, a limitation of these tools is their general-purpose nature, which may not fully account for the unique and dynamic challenges faced during a penetration test. Customization of these tools to specific pentesting environments or frameworks could greatly enhance their performance.

In addition to supporting professional pentesters, these tools can also serve as valuable educational aids, providing novice pentesters with examples of attack scenarios, mitigation strategies, and detailed reports for training purposes. To maximize the effectiveness of GenAI tools, future work should explore the integration of these tools with established pentesting frameworks and technologies. By combining the strengths of GenAI with proven tools, the pentesting process could be significantly enhanced, improving both the speed and accuracy of assessments. Another critical factor influencing outcomes is the human skill involved in crafting prompts. The effectiveness of LLMs in pentesting depends significantly on how precisely the pentester formulates queries and structures context. As such, the productivity gains offered by GenAI tools are partially contingent on the pentester’s experience and knowledge in both cybersecurity and prompt engineering. It is also important to point out that using these GenAI tools in real-world pentesting poses legal and ethical risks, such as unsafe command suggestions, data privacy violations, and inaccurate outputs. Pentesters should validate AI-generated commands, avoid sharing sensitive data, maintain human oversight, and document AI usage. Clearer guidelines are needed as the technology evolves.

As future work, several promising directions for future research and development emerge. First, conducting more extensive evaluations of these GenAI tools in diverse pentesting scenarios could provide deeper insights into their applicability across different environments. Second, as specialized GenAI tools for pentesting continue to evolve, future studies should investigate whether these tools are sufficiently robust to operate independently or if they would be better integrated with existing solutions. Finally, an interesting avenue for exploration would involve training a GenAI model specifically tailored for pentesting, drawing from pentesting reports, attack models, vulnerability databases, and other pertinent sources of information to optimize its capabilities.

\backmatter

\section*{CRediT authorship contribution statement}
\textbf{Antonio López Martínez}: Writing – review \& editing, Writing – original draft, Visualization, Validation, Investigation, Conceptualization. \textbf{Alejandro Cano}: Writing – review \& editing, Supervision, Validation, Investigation, Conceptualization. \textbf{Antonio Ruiz-Martínez}: Writing – review \& editing, Writing – original draft, Visualization, Validation, Investigation, Funding acquisition, Conceptualization.

\section*{Acknowledgements}

The research was supported by DOSS, which has received funding from the European Union’s Horizon Europe research and innovation programme under grant agreement No 101120270, and CERBERUS, TSI-063000-2021-36, 44, 45, 62 (CERBERUS), funded by “Ministerio de Asuntos Económicos y Transformación Digital“ (MAETD)’s 2021 UNICO I+D program.

\section*{Declaration of competing interest}
The authors declare that they have no known competing financial interests or personal relationships that could have appeared to influence the work reported in this paper.

\section*{Declaration of generative AI and AI-assisted technologies in the writing process}

During the preparation of this work the author(s) used ChatGPT and Claude 3.5 Sonnet in order to improve writing and readability. After using this tool/service, the author(s) reviewed and edited the content as needed and take(s) full responsibility for the content of the publication.

\bibliography{sn-article}

\end{document}